\documentclass[12pt]{article}
 \usepackage{graphicx}
 \usepackage{amssymb,amsmath}
\begin{document}

\title{Interaction between antiprotonic helium ion and He atom: Potential Energy Surface}
\author{S.N. Yudin$^1$,  I.V. Bodrenko$^2$,  and  G.Ya. Korenman$^1$ \\
$^1$\emph{\small D.V.Skobeltsyn Institute of Nuclear Physics,
 M.V. Lomonosov Moscow State University,}\\ \emph{\small Moscow 119991,  Russia}\\
{\small E-mail: korenman@nucl-th.sinp.msu.ru}\\
$^1$\emph{\small Department of Physics, University of Cagliari, 
Cittadella Universitaria di Monserrato,}
\\  \emph{\small S.P.8−km 0.700, 09042 Monserrato, Cagliari, Italy}}

\date{}
\maketitle

\begin{abstract}
Potential Energy Surface for the ($\mathrm{\bar{p}} - \mathrm{He}^{2+} -
\mathrm{He}$) system is calculated in the framework of the restricted (singlet
spin state) Hartree-Fock method with subsequent account of the electronic
correlations within the second order perturbation method (MP2). The geometry of
heavy particles is described in variables of distance $r$ from nucleus $a$ to
antiproton, distance $R$ from the center of mass of the $(\bar{p} - a)$ pair to
$\mathrm{He}$ atom containing nucleus $b$, and an angle $\theta$  between
$\mathbf{r}$ and $\mathbf{R}$. The potential $V(R,r,\cos\theta)$ of the interaction
between $\mathrm{He}$ atom and a $\bar{p} - a$ subsystem involves a total
energy of two electrons in the field of three heavy particles and
Coulomb interactions of the $b$  nucleus with antiproton and the $a$ nucleus. The
expansion of this potential in terms of Legendre polynomials
$\mathrm{P}_k(\cos\theta)$ is obtained. Matrices of 
the multipole terms $V^k(r,R)$ ($k=0,1,2$) are obtained  in the basis of antiprotonic helium ion states.
The results are compared with the model potential that was used earlier in the
calculations of collisional Stark transitions of $(\mathrm{\bar{p}He}^{2+})$
ion. Total cross sections of collisional Stark transitions obtained with the PES potentials exceed the model results by 15 - 20\% at $E \leq 12$ K. 
\end{abstract}

\section{Introduction} \label{sec1}
The experimental discovery and investigation of the metastable antiprotonic
states in helium have opened a new chapter in the study of antiprotonic atoms
(see comprehensive reviews \cite{ref1, ref2}). In addition to the fundamental
properties of the antiproton, these experiments give interesting insights
into the interaction of antiprotonic helium with ordinary atoms and molecules.
The data stimulated many theoretical papers on the mechanisms of antiprotonic
helium formation, collisional quenching of the metastable states, effects
of collisional (density) shift and broadening of E1-spectral lines, \textit{etc}.
The detailed study of metastable states of antiprotonic helium by the ASACUSA
collaboration at the AD beam (CERN) brings further data that pose specific
problems in the theory of collisional effects at low temperature ($\sim 10$ K).
Thus, the measurements of hyperfine splitting of antiprotonic helium levels by
a triple laser-microwave-laser resonance method \cite{ref3, ref4, ref5} and
their interpretation require calculations of the rates of collisional
transition between HFS sublevels as well as the density shifts and broadenings
of the microwave M1 spectral lines \cite{ref6, ref7}.

Among many experimental results obtained by ASACUSA collaboration, there is one
very specific: the first observation of cold long-lived antiprotonic helium
ions $(\bar{p}\mathrm{He}^{2+})$ in the states with definite quantum numbers
($n=28\, - \,32, \,l=n-1$) \cite{ref8}. It was observed that these states have
lifetimes $\tau\sim 100$ ns against annihilation, and rates of annihilation
$\lambda=\tau^{-1}$ increased roughly as a linear function of target density at
$\rho<(5\, - \,10)\cdot 10^{17}$ cm$^{-3}$ and temperature $T\simeq 10$ K. The
antiproton annihilation occurs in the states with small angular momentum, therefore
the observed lifetime has to be attributed to the time of
transitions from the initial state ($l_i\simeq 30$) to the $S,\,P,\,D$ states.
Radiative transitions from the circular orbits ($l=n-1$) at large $n$ are
strongly suppressed, therefore a main contribution to the effective rate of
annihilation comes from collisional transitions. In this case the observed
values of $\lambda$ (at the given $\rho$ and $T$) require to have an effective
cross section for the initial states $\sigma_q\sim (4\, - \,10)\cdot 10^{-15}$
cm$^2$. This value is an order of magnitude larger than previous theoretical
cross sections for collisional Stark transitions \cite{ref9, ref10, ref11}
relevant to higher kinetic energy corresponding to $T\sim 10^4$ K. In addition
to huge quenching cross sections, the experiment revealed an unexpected
$n$-dependence and isotope effect ($A=3,\,4$) for effective annihilation rates. These
results also cannot be explained by earlier considerations of collisional
processes of $(\bar{p}\mathrm{He}^{2+})$.

\begin{figure}[thb]
\includegraphics[width=0.7\textwidth]{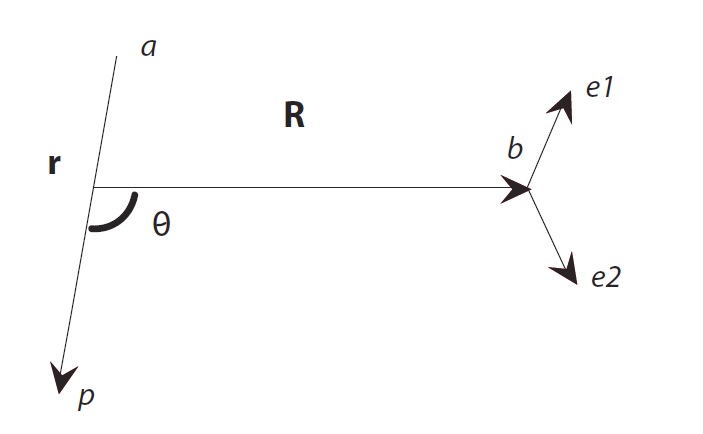}
    \caption{Jacobi coordinates of heavy particles}
    \label{fig1}
\end{figure}

In the papers \cite{ref12, ref13} collisions of cold $(\bar{p}\mathrm{He}^{2+})_{nl}$ ion with 
surrounding helium atoms at $T\sim 10$ K were considered  in the framework of close coupling 
approach with account for the states with all angular momenta $l$ from $n-1$ to 0 at the fixed $n$. 
A model interaction between two colliding subsystems was taken as
\begin{equation}
 V(R,r,\cos\theta)  = V_0(R) + V_1(R) P_1(\cos\theta) + V_2(R)  P_2(\cos\theta) \label{eq1} , 
\end{equation}
where $V_0(R)$ is an adiabatic potential of the interaction between the ionic charge (+1) and He 
atom, $R$ and $r$ are distances from the ion center of mass  to He atom and from $\mathrm{He}
^{2+}$ nucleus to $\bar{p}$, respectively, and $\theta$ is the angle between vectors $\mathbf{R}$ 
and $\mathbf{r}$ (see Fig.\ref{fig1}). Second and third terms in Eq.\eqref{eq1} take into account interactions of 
electric dipole and quadrupole momenta of the ion with He atom polarized by the unit ionic charge, so
\begin{align}
 V_1(R) &= \xi_1\, rV_0^\prime(R)	\label{eq2}, \\
V_2(R)  &= \frac{1}{2}\xi_2 r^2  R\left(R^{-1}V_0^{\prime}(R)\right)^{\prime}, \label{eq3}
\end{align}
  Factors $\xi_1=(M_A+2m_p)/(M_A+m_p)$ and $\xi_2=(2m_p^2 - M_A^2
)/(M_A+m_p)^2$ take into account a displacement of a center of charge with respect to center of 
mass in the $\bar{p}A^{2+}$ system and, for $\xi_2$, a difference of two masses too. A 
correction of order $r^2$ to the scalar term in \eqref{eq1} is omitted.
 The dipole and quadrupole terms in \eqref{eq1} can mix $nl$ states of the ion and lead to Stark 
transition. Induced annihilation due to mixing with $S$ and $P$ states during collisions was also 
taken into account. The potential $V_0(R)$ is, in fact, the same as the adiabatic potential 
of the interaction between proton and He atom. It was approximated by the sum of the Morse 
potential $V_M(R)$ and the long-range polarization potential cut off at an intermediate distance,
\begin{align}
V_0(R) & = V_M(R)+ V_p(R), \label{eq4} \\
V_m(R) & = D_0 \left\{\exp[-2\beta(R-R_e)] -
            2\exp[-\beta(R-R_e)]\right\}, \label{eq5} \\
V_p(R) & = -\frac{\alpha}{2R^4} \left\{1 -
\exp\left[-\gamma(R-R_e)^4\right]\right\}\theta(R-R_e) , \label{eq6}
\end{align}
with the parameters\cite{ref14,ref15} $\alpha=1.383$, $D_0=0.075$, $R_e =1.46$, $\beta =1.65$,
$\gamma=0.005$. Here and further in this work we use atomic units ($\hbar=e=m_e=1$) unless 
otherwise specified.

This model gives cross sections of Stark transitions from circular orbits with $n\sim 30$ at $E=10$
K that are of order or greater than effective quenching cross section $\sigma_q$. Stark transition 
rates averaged over thermal motion correlate similarly with the observed effective annihilation rates. Thus, 
the model allows to understand qualitatively the observed data. Moreover, it gives also some isotope effect with correct sign and an increasing of the rates with $n$. However, if to take into account a whole cascade of Stark and radiative 
transitions from the initial circular orbit up to annihilation from $S$ and $P$ states, it turns out that 
the obtained Stark cross sections have not enough high values.

A possible improvement of the theory can be to use an \textit{ab initio} Potential Energy Surface (PES) for three heavy particle in the $(\mathrm{\bar{p}} - \mathrm{He}^{2+} - \mathrm{He})$ system 
instead of the model potential \eqref{eq1} - \eqref{eq4}. In this paper we calculate
the potential energy surface for the mentioned system in the framework of the restricted (singlet
spin state) Hartree-Fock method with subsequent account of the electronic correlations within the 
second order perturbation method (MP2). There are very few publications on \textit{ab initio} 
PES  calculations for systems involving antiproton and two or three nuclei. As far as we know, 
potential energy surfaces for the $(\bar{p}\mathrm{He}^+) - \mathrm{He}$ and $(\bar{p}\mathrm{He}^+) - \mathrm{H}_2$ systems only 
were considered in the literature (see papers \cite{ref16, ref17} and \cite{ref18}, respectively). 
A system under consideration in this paper $(\mathrm{\bar{p}} - \mathrm{He}^{2+} - \mathrm{He})$  seems to 
be more simple because it has only two electrons instead of three electrons in the above mentioned 
systems. On the other hand, our system has nonzero total charge that leads to a long-range 
polarization interaction between subsystem 'antiprotonic ion - He atom'. So far as we know the
publications on potential energy surface for the $(\mathrm{\bar{p}} - \mathrm{He}^{2+} - \mathrm{He})$ 
system are absent in the literature.

The paper is organized as follows. In Sec. \ref{sec2} we introduce notations, outline briefly the 
used method of PES calculations, and define the Jacobi coordinates grid points for the calculation. 
The results for a general 3D view of the potential and more detailed dependencies of the PES on 
the variables $\cos\theta$, $r$ and $R$ as well as the results for calculations of the multipole 
expansion of interaction between atom $\mathrm{He}$ and a $\bar{p} -\mathrm{He}^{2+}$ 
subsystem are presented in Sec. \ref{sec3}. Matrices of the multipole terms in the basis of $(\bar{p}
\mathrm{He}^{2+})_{nl}$ states are considered and compared with the model results in Sec. \ref{sec4}. 
With these matrices we have calculated total cross sections of collisional Stark transitions. The cross sections obtained with the PES potentials exceed the model results by 15 - 20\% at $E \leq 12$ K. 
A general conclusion is given in Sec. \ref{sec5}.
\section{Method of calculation of the potential energy surface} \label{sec2}

The system under consideration ($\mathrm{\bar{p}} - \mathrm{He}^{2+} -
\mathrm{He}$) consists of two electrons and three heavy particles. Let
$m_\mathrm{\bar{p}}$, $M_a$, $M_b$ and $\mathbf{r}_\mathrm{\bar{p}}$,
$\mathbf{R}_a$, $\mathbf{R}_b$ be masses and coordinates of antiproton and two
helium nuclei, respectively. The indexes $a$ and $b$ refer to the nuclei in $(\mathrm{\bar{p}} 
\mathrm{He}^{2+})$ and $\mathrm{He}$ atom subsystems, respectively. The c.m. coordinates 
of He atom coincide with the coordinates of corresponding nucleus, $\mathbf{R}_{\mathrm{He}} 
\simeq \mathbf{R}_b$ with an accuracy of order $m_e/M_b$. The configuration of the heavy 
particles is defined by the Jacobi coordinates $\mathbf{r}=\mathbf{r}_\mathrm{\bar{p}}-\mathbf{
R}_a$ and 
$\mathbf{R}= \mathbf{R}_b - (\lambda \mathbf{r}_\mathrm{\bar{p}} + \nu \mathbf{R}_a)$, 
where 
$\lambda = M_a/(M_a + m_\mathrm{\bar{p}})$, $\nu = m_\mathrm{\bar{p}}/(M_a + m_
\mathrm{\bar{p}})$.
The heavy particles in the problem under consideration are very slow that allows to use adiabatic 
approximation. Then an interaction potential between two subsystem, ($\mathrm{\bar{p}} - 
\mathrm{He}^{2+}$) and $\mathrm{He}$ atom, can be presented as
\begin{equation} \label{eq7}
V(r, R, \cos\theta) = \frac{4}{|\mathbf{R} + \nu \mathbf{r}|} - \frac{2}{|\mathbf{R} - \lambda \mathbf{r}|} + E_e(r, R, \cos\theta) - E_\mathrm{He} ,
\end{equation}
where $\theta$ is the angle between the vectors $\mathbf{r}$ and $\mathbf{R}$, 
$E_e(r, R, \cos\theta)$ is the total energy of two electrons in the field of three heavy particle ($a,\,b
$ and $\bar{p}$), and $E_\mathrm{He}$ is the ground state energy of the isolated $\mathrm{He}$
 atom. The first two terms in \eqref{eq7} take into account Coulomb interaction between the $(a - \bar{p})$ 
subsystem and the nucleus $b$. Coulomb interaction between the nucleus $a$  and antiproton  ($-2
/r$) is absent in \eqref{eq7} because it has to be included into inner energy of the 
$(\mathrm{\bar{p}} \mathrm{He}^{2+})$ ion.
  The potential \eqref{eq7} at $r\ll R$ reduces to the form of Eq. \eqref{eq1}.
  
The total energy of two electrons in the field of three heavy particle $E_e(r, R, \cos\theta)$ as well 
as the energy $E_\mathrm{He}$ of the helium atom in the ground state were calculated in the 
framework of the restricted (singlet spin state) Hartree-Fock method with subsequent account of the 
electronic correlations within the second order perturbation method (MP2). We used the augmented 
correlation-consistent polarized valence sextuple zeta aug-cc-pV6Z and, for a comparison, quintuple 
zeta aug-cc-pV5Z molecular basis sets \cite{ref19,ref20} in the Cartesian form. The main part of 
the calculations involves the orbitals centered on the helium nucleus $b$ as well as the orbitals 
centered on an effective 'nucleus' with the unit positive charge placed in the $(\mathrm{\bar{p}} 
\mathrm{He}^{2+})$ center of mass. The parameters of the basis set for the latter center were 
taken as for hydrogen.  Additional calculations with the orbitals centered only on the helium 
nucleus $b$ were also done showing a reasonable results at $R\gg r$.

Parameters of the basis sets were obtained from the Extensible Computational Chemistry 
Environment Basis Set Database \cite{ref21}. To overcome the basis set superposition error (BSSE), 
at every configuration, the calculations were performed with the same basis set for both the 
helium atom and the system of two electrons in the field of three heavy particles.

For the calculations, we used the original program taking into account a negative charge of one of 
heavy particles (author I.V. Bodrenko). The program employs the RI ('resolution of identity') 
method for the electron-electron interaction integrals and allows to perform the Hartree-Fock based 
calculations for large systems or for a number of configurations with moderate computational 
resources (see \cite{ref22,ref23} for details). 

The potential energy was calculated at the geometry configurations corresponding to the following
regular grid in the Jacobi coordinates (initial value, step, final value):
$r = 0.1, 0.04, 0.9$, $R = 0.5 , 0.1, 10$,  $\cos\theta = -1 , 0.1, +1$; totally 42336 
configurations. At the each grid point the Jacobi coordinates were transformed to the Cartesian 
ones by using the following masses of the particles (in a.m.u.): $m_{\bar{p}} = 1.0073$; $M_{^4 
\mathrm{He}} = 4.0015$, $M_{^3\mathrm{He}} = 3.0149$.
The calculations with two basis sets (the quintuple zeta aug-cc-pV5Z and sextuple zeta aug-cc-pV6Z)
 give the effectively indistinguishable values of $V(r, R, \cos\theta)$ in the mentioned region, 
therefore we will show below the results  only for the more extended basis set (pV6Z).

\section{Results for the Potential Energy Surface} \label{sec3}

\begin{figure}[thb]
\centering
    \includegraphics[width=0.32\textwidth]{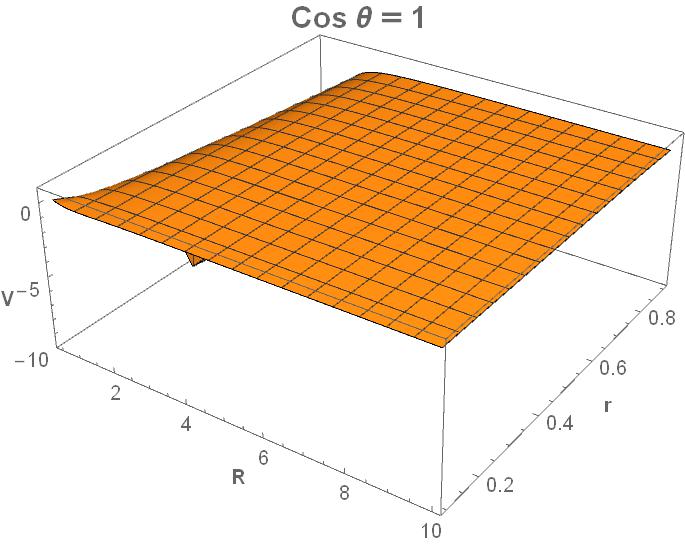}
		\put(-90,140){(a)}
		\includegraphics[width=0.32\textwidth]{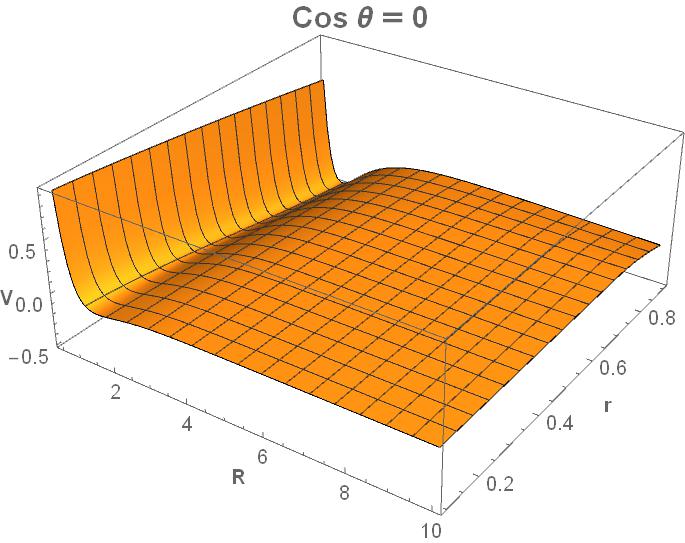}
 \put(-90,140){(b)}
\includegraphics[width=0.32\textwidth]{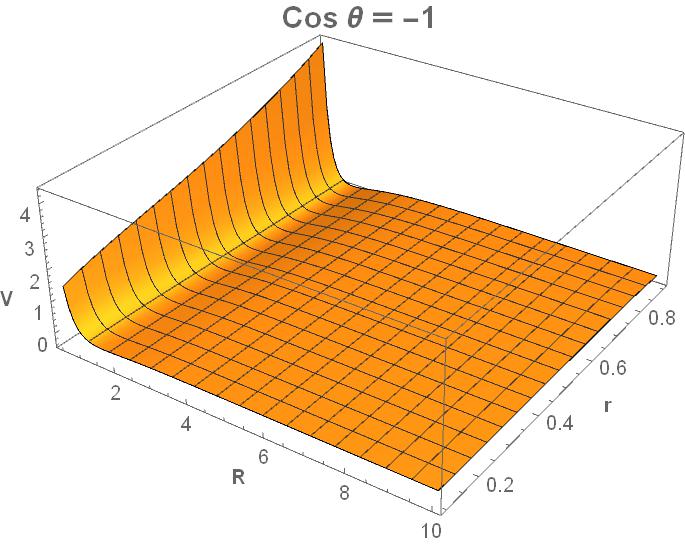}
\put(-90,140){(c)} 
    \caption{Potential energy surface for the system $\bar{p}-\mathrm{He}^{2+}-\mathrm{He}$ at 
(a) $\cos\theta=1$ (antiproton between two nuclei), (b) $\cos\theta=0$ and (c) $\cos\theta=-1$.}
    \label{fig2}
\end{figure}

A general view of the potential energy surface $V(r, R, \cos\theta)$ depending on $r$ and $R$ is 
shown in Fig. \ref{fig2} for three different angles between $\mathbf{r}$ and $\mathbf{R}$ ($
\cos\theta$=+1 corresponds to antiproton position between two nuclei). The surface is very smooth except near the points $|\mathbf{R} + \nu \mathbf{r}|=0$ and $|\mathbf{R} - \lambda \mathbf{r}| = 0$, where one of the Coulomb terms in Eq. \eqref{eq7} tend to infinity.  

\begin{figure}[thb]  
\centering
\begin{minipage}{17pc}
\vspace{-4mm}
    \includegraphics[width=\textwidth]{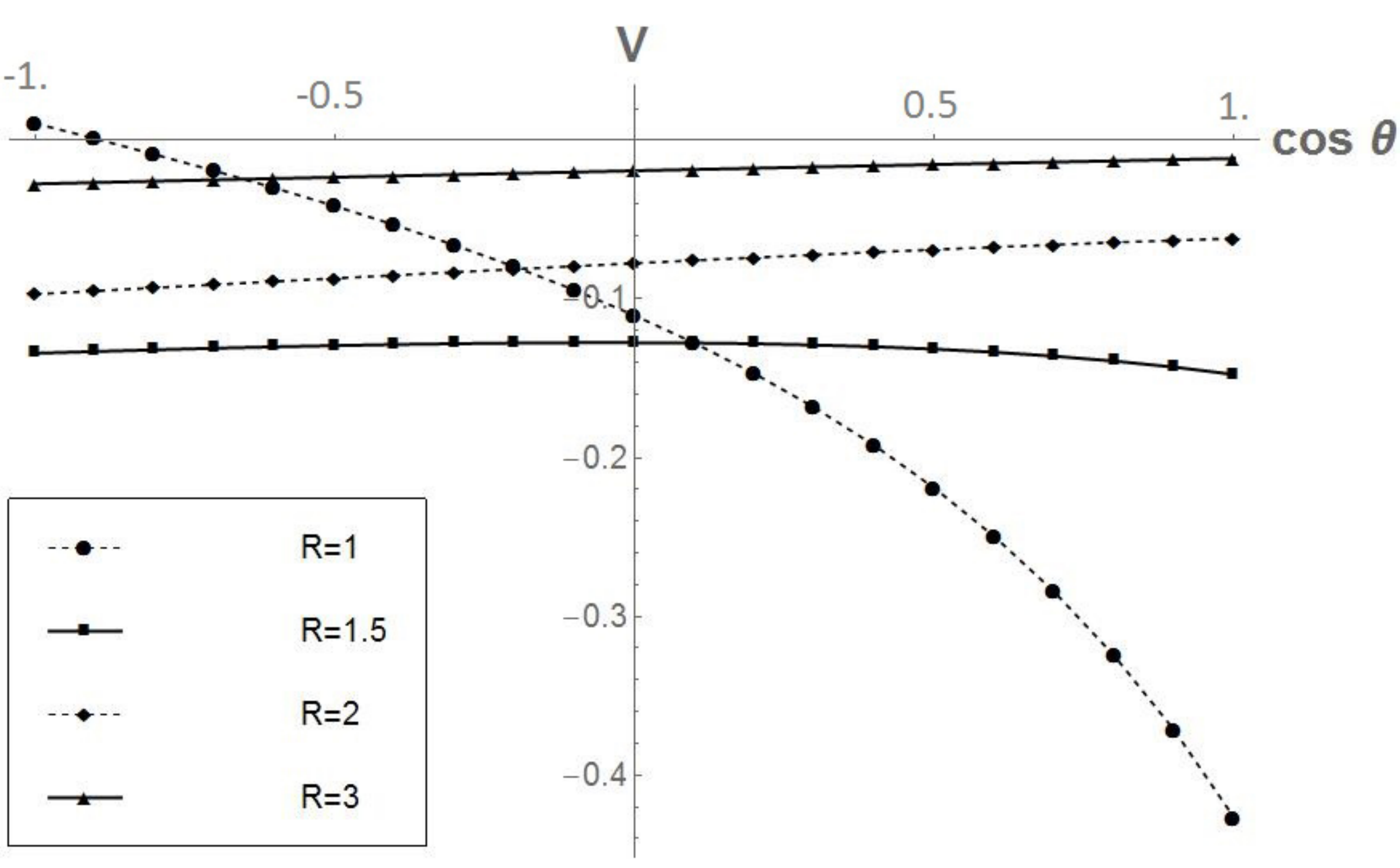}
		\put(-90,135){(r=0.3)}
\end{minipage}\hspace{2pc}		
\begin{minipage}{17pc}		
		\includegraphics[width=\textwidth]{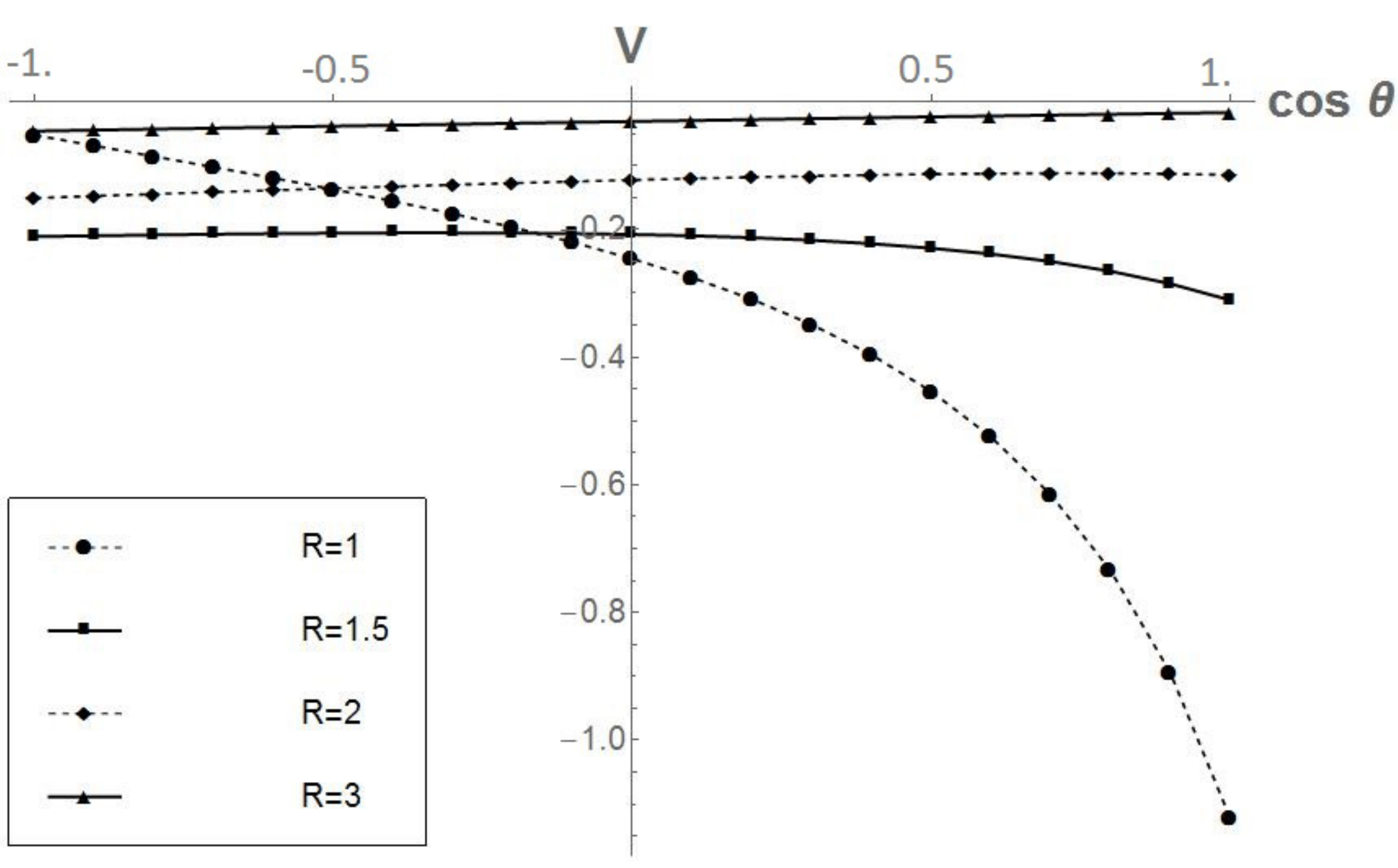}
 \put(-90,140){(r=0.5)}
\end{minipage}
    \caption{Dependence of PES on $\cos\theta$ for the system $\bar{p}-\mathrm{He}^{2+}-
\mathrm{He}$ at $r=0.3$ (left part) and $r=0.5$ (right part) for different values of $R$.}
    \label{fig3}
\end{figure}

The potential as a function of $\cos\theta$ is shown in Fig. \ref{fig3} for several values of $R$ at $r
=0.3$ (left part of the figure), 0.5 (right part). These values of $r$ are about the mean radii of 
antiprotonic helium ion in the states $n=30, l=n-1$ and $l=0$, respectively.
It is seen from the Figure that a dependence of the potential  on $\cos\theta$ is rather weak in the 
region $r\leqslant 0.6$, $R\geqslant 1$. Therefore the expansion of the potential energy surface in 
Legendre polynomials
\begin{equation} \label{eq8}
V(r,R,\cos\theta) = \sum^\infty_{k=0} V^k(r,R) P_k (\cos\theta)
\end{equation}
can be restricted by lowest multipoles ($k=0,\,1,\,2$).

\begin{figure}[thb]
\centering
    \includegraphics[width=0.32\textwidth]{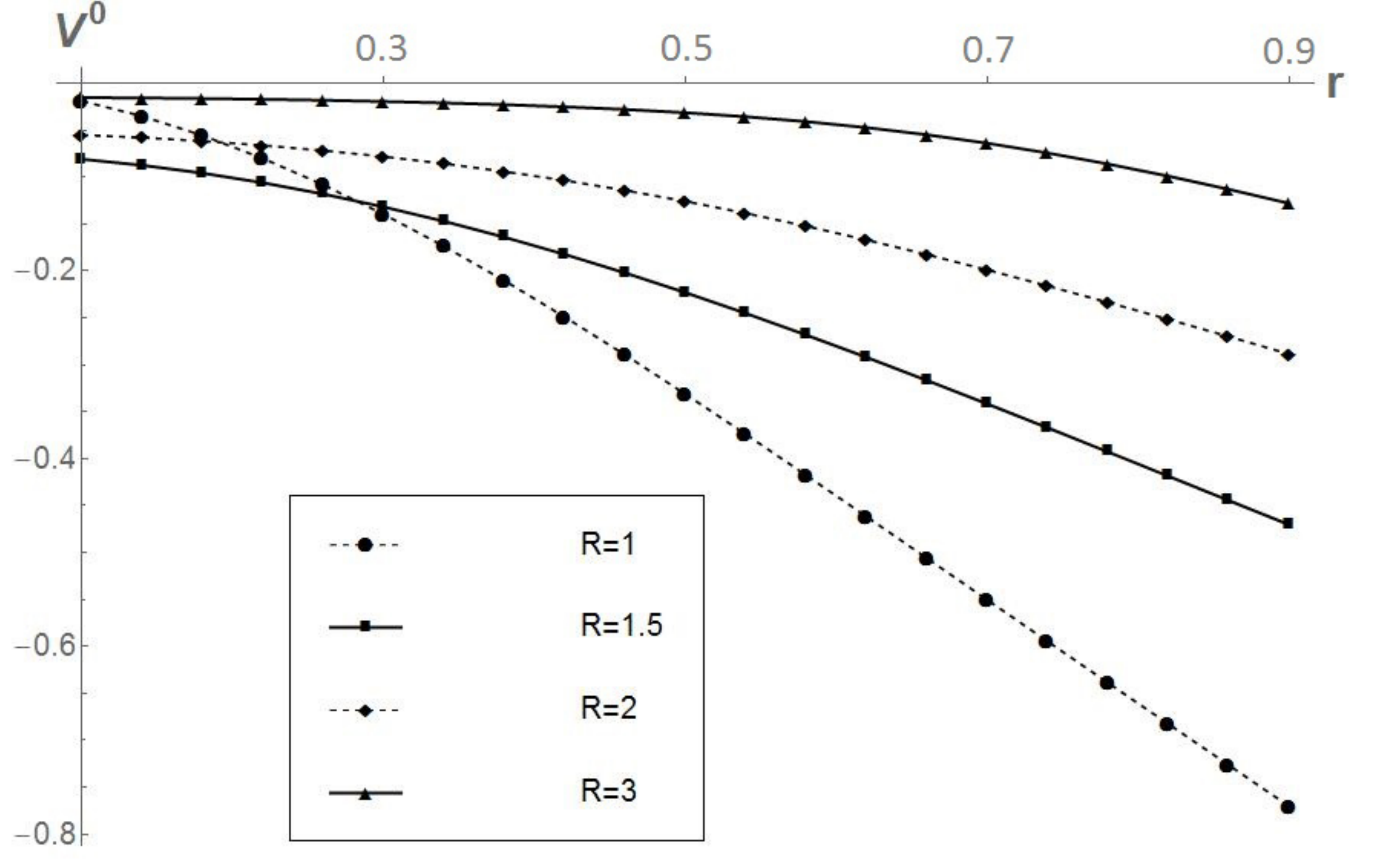}
		\put(-90,105){(a)}
		\includegraphics[width=0.32\textwidth]{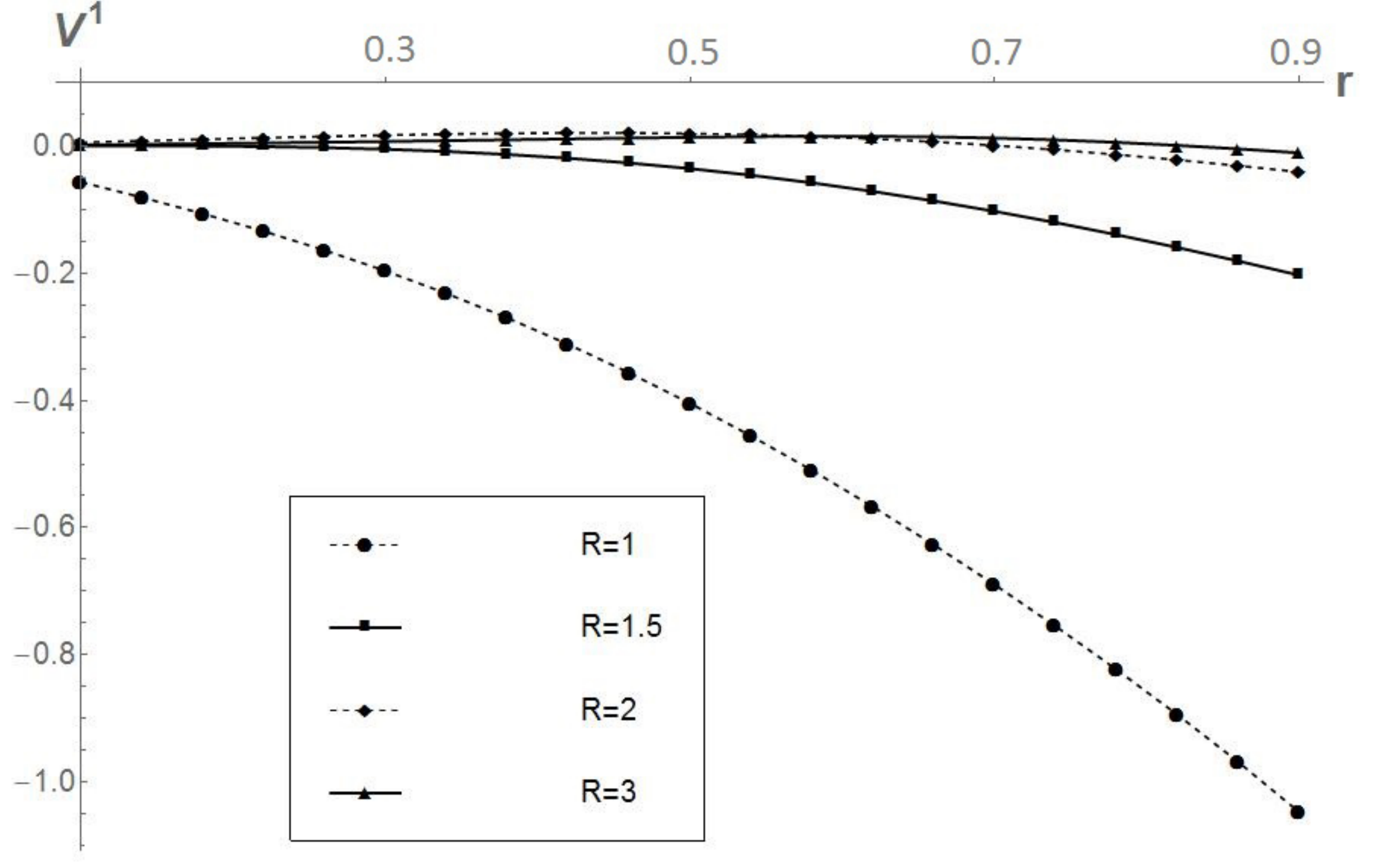}
 \put(-90,105){(b)}
\includegraphics[width=0.32\textwidth]{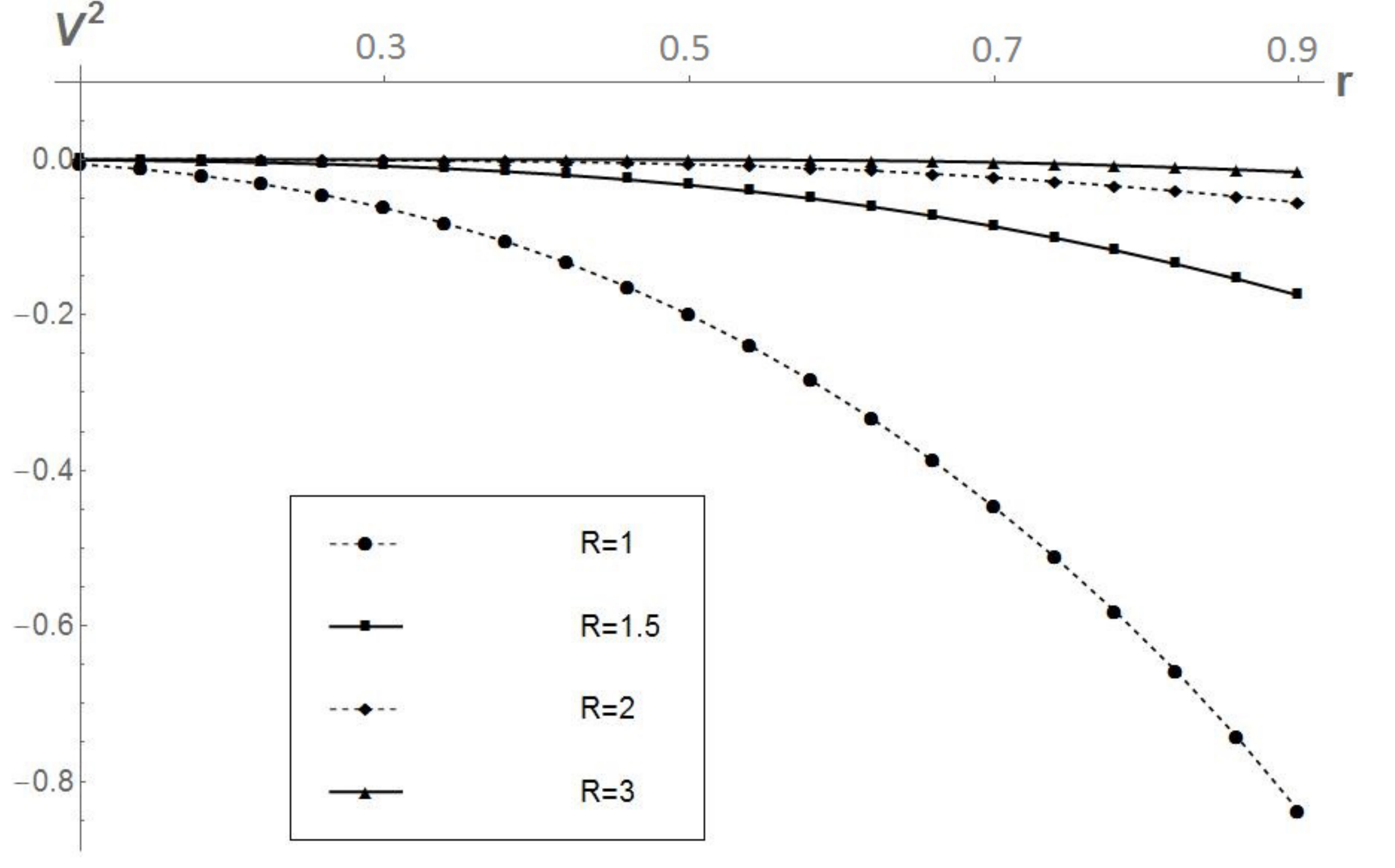}
\put(-90,105){(c)} 
    \caption{Dependence of scalar (a), dipole (b) and quadruple (c) terms of the multipole expansion \eqref{eq8} on $r$ at several $R$.}
		\label{fig4}
	\end{figure}

\begin{figure}[thb]
\centering
    \includegraphics[width=0.32\textwidth]{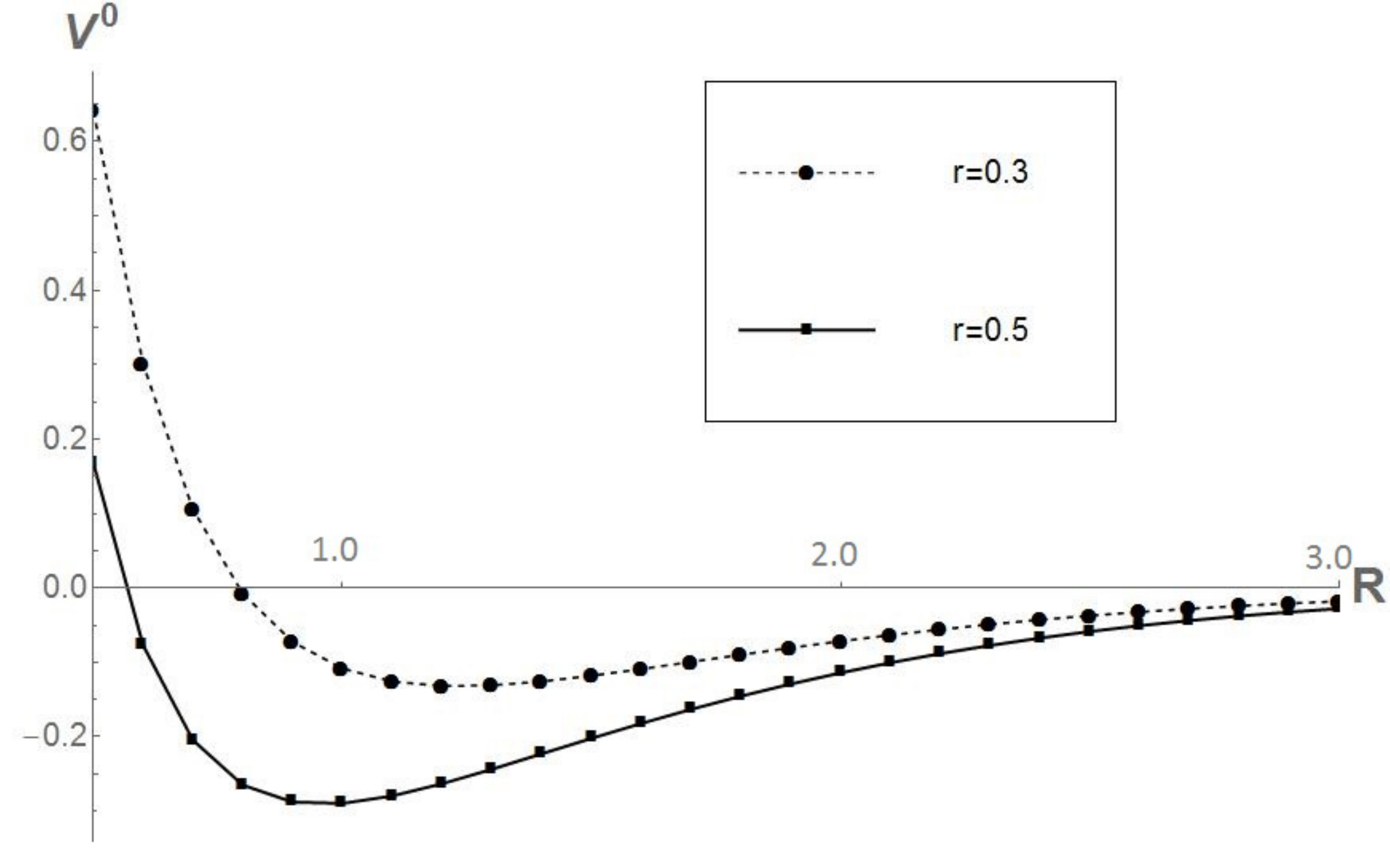}
		\put(-90,105){(a)}
		\includegraphics[width=0.32\textwidth]{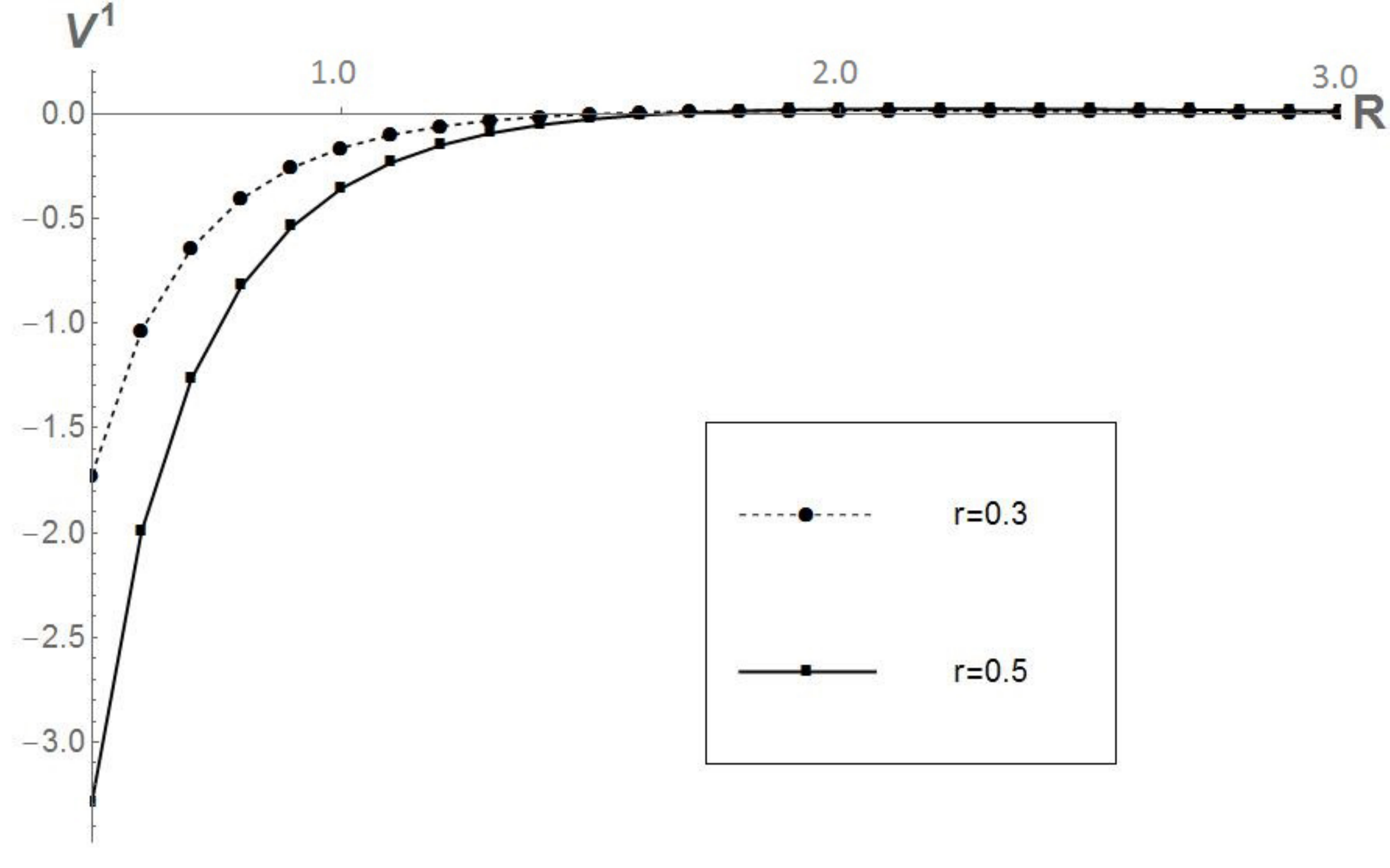}
 \put(-90,105){(b)}
\includegraphics[width=0.32\textwidth]{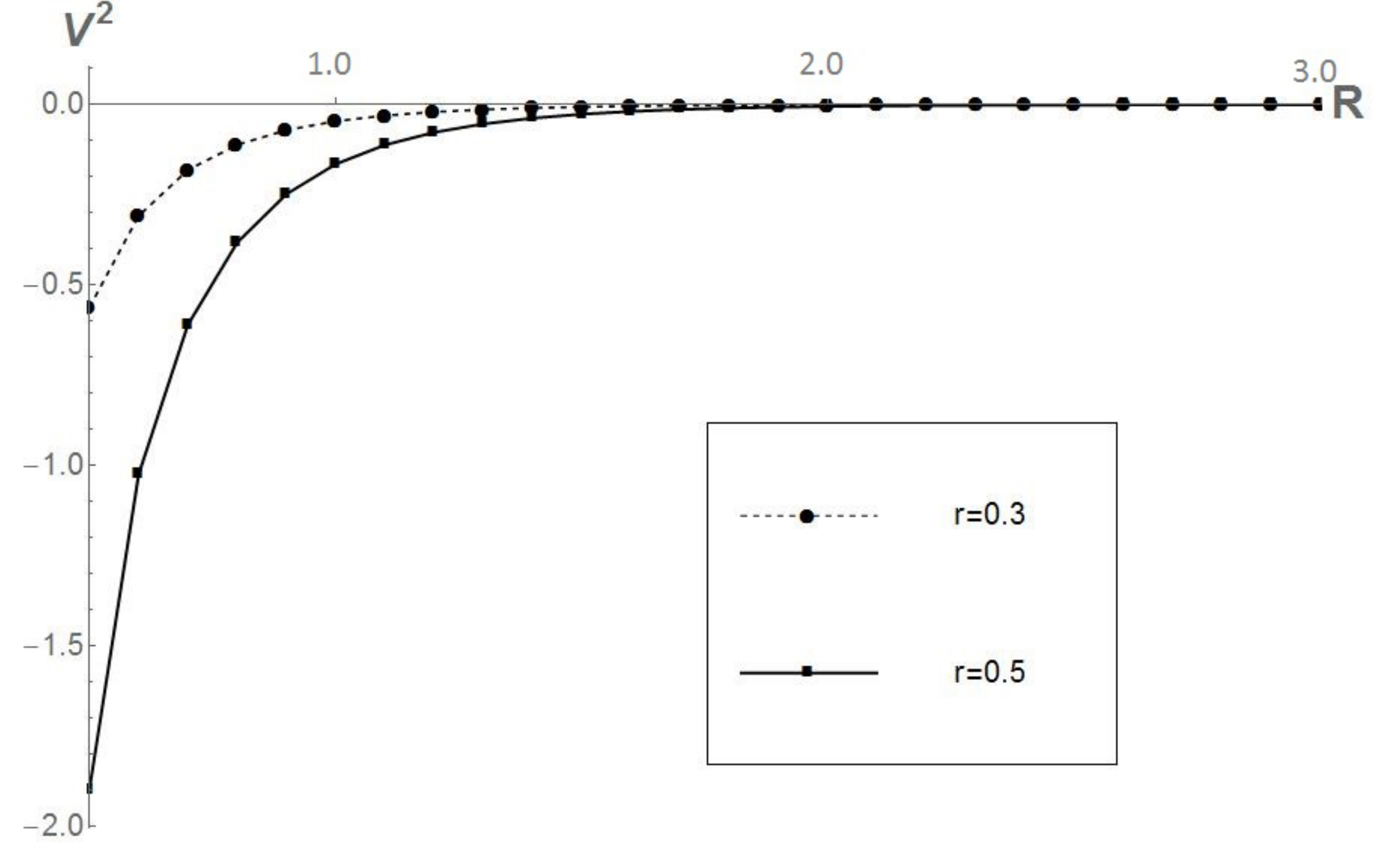}
\put(-90,105){(c)} 
    \caption{Dependence on $R$ of scalar (a), dipole (b) and quadruple (c) terms of the multipole expansion \eqref{eq6} at $r=0.3$ and 0.5.}
		\label{fig5}
	\end{figure}

Dependencies of the multipole terms $V^0(r,R)$, $V^1(r,R)$ and $V^2(r,R)$ on $r$ and $R$ are 
shown in Figs. \ref{fig4} and \ref{fig5}. General forms of  these terms are qualitatively similar to the 
corresponding model terms $V_0(R)$, $V_1(r,R)$ and $V_2(r,R)$, which are given by Eqs. \eqref{eq2} - \eqref{eq6}. However the \textit{ab initio} values may be quantitatively different from the 
model ones. It can lead to some distinctions in matrix elements (see next Section).
\section{Matrix of the potential energy surface in the basis of antiprotonic helium ion states} \label{sec4}
\begin{figure}[thb]  
\centering
    \includegraphics[width=0.48\textwidth]{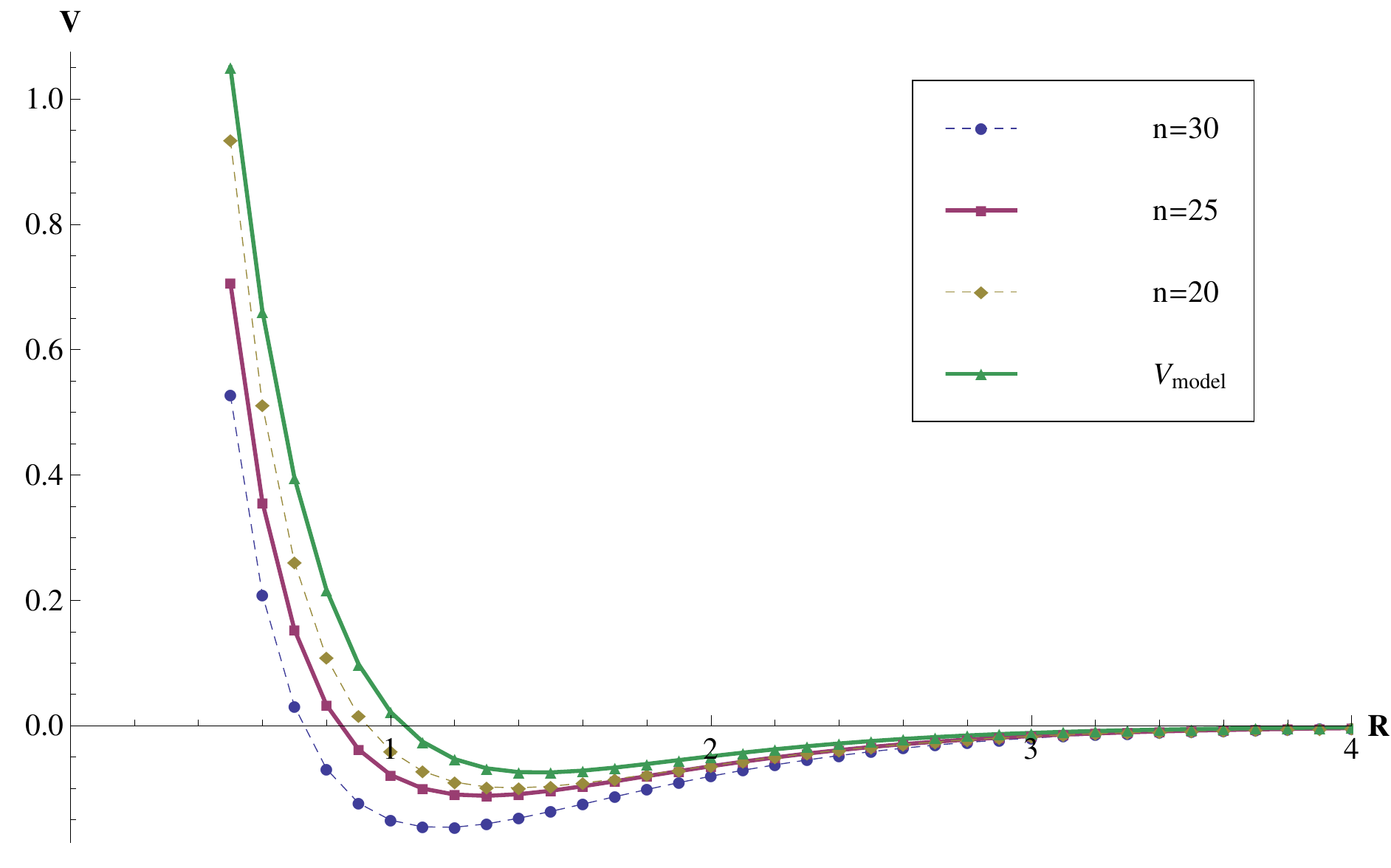}
		\put(-120,140){(a)}
		\includegraphics[width=0.48\textwidth]{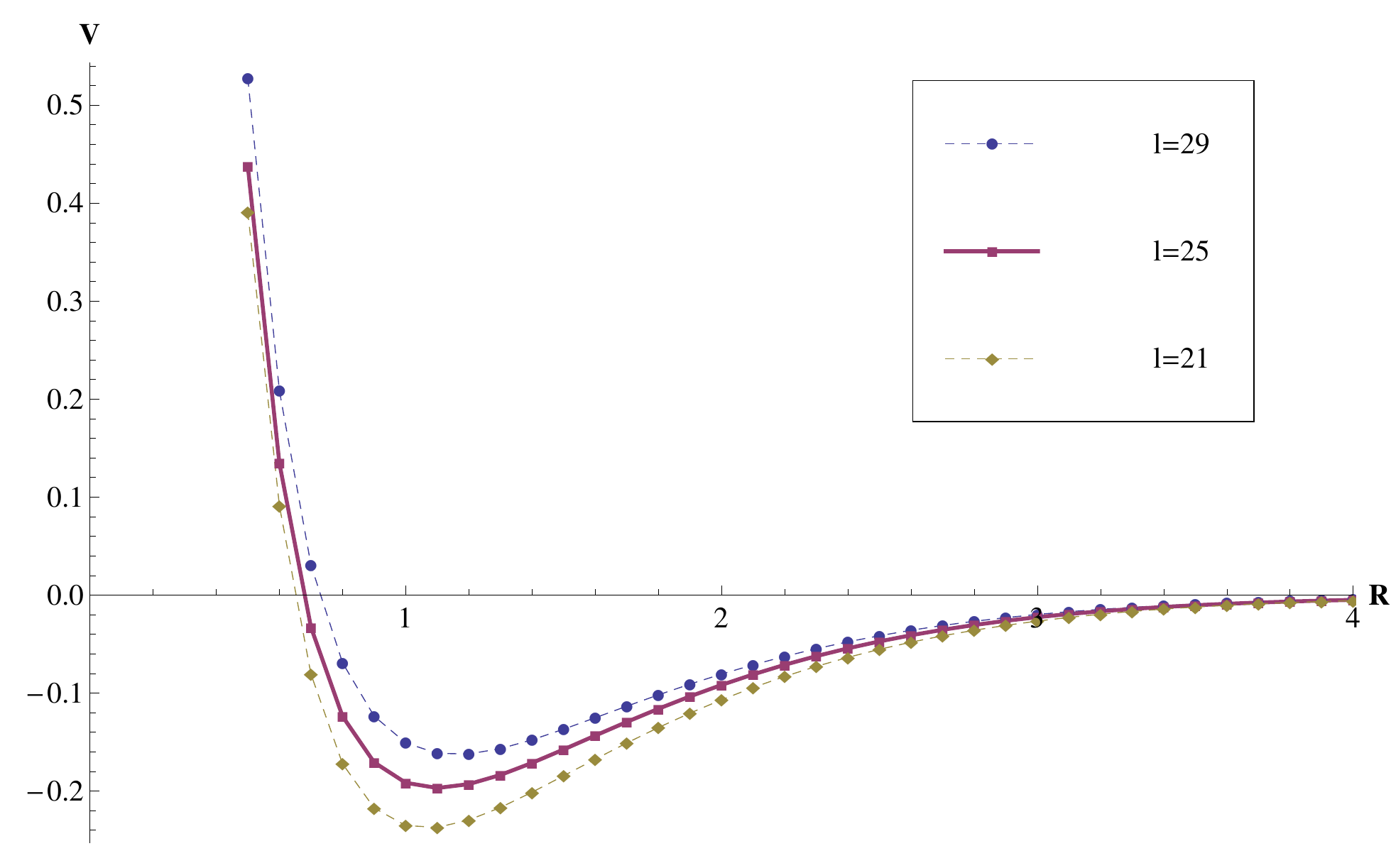}
 \put(-120,140){(b)}
\caption{Dependence of $V^0_{nl}(R)$ on $R$ (a) for circular orbits ($l=n-1$) at different $n$ and (b) for different $l$ at $n=30$. The model potential $V_0(R)$ is also shown on the part (a).}
    \label{fig6}
\end{figure}

Potential energy surface considered in the previous section can be used, instead of the model \eqref{eq1} - \eqref{eq6}, within a quantum close-coupling approach to the collisions of $(\bar{p}\mathrm{He}^{2+})$ ion with $\mathrm{He}$ atoms. 
With this in mind, we consider the potential matrix in the basis of antiprotonic helium ion  wave functions
\begin{equation} \label{eq9}
\langle nl'm'|V(r,R,\cos\theta)|nlm\rangle =\delta_{ll'}\delta_{mm'}V^0_{nl}(R) + \sum^\infty_{k=1} V^k_{nl',nl}(R)\cdot \langle Y_{l'm'}(\Omega_r)|P_k (\cos\theta)|Y_{lm}(\Omega_r)\rangle ,
\end{equation}
where 
 \begin{align}
 V^0_{nl}(R) & = V^0_{nl,nl}(R), \label{eq10} \\
 V^k_{nl',nl}(R) & = \int_0^\infty R_{nl'}(r)  V^k(r,R) R_{nl}(r) r^2 dr , \label{eq11}\\
 \langle Y_{l'm'}(\Omega_r)|P_k (\cos\theta)|Y_{lm}(\Omega_r)\rangle &=\langle l0\,k0|l'0\rangle \sqrt{4\pi/(2k+1)}
\sum_q Y_{kq}^*(\Omega_R)\cdot \langle lm\,kq|l'm'\rangle\ , \label{eq12}
 \end{align}
$R_{nl}(r)$ are the radial hydrogen-like wave functions of antiprotonic helium ion, and $Y_{lm}(\Omega_r)$ are the ordinary spherical functions. 
Diagonal matrix elements \eqref{eq10} have the same sense as monopole term $V_0(R)$ in the model 
potential \eqref{eq1}, but the formers are calculated from \textit{ab initio} PES and depends on 
the $(\bar{p}\mathrm{He}^{2+})_{nl}$ ion state. The calculated potentials $V^0_{nl}(R)$ 
for different $n$ and $l$ are shown in Fig. \ref{fig6}. For the comparison, the model potential $V_
0(R)$ is also shown on the figure, part (a). It is seen from the figure that all calculated curves are qualitatively similar to the model potential, 
however they have more deep minima that are shifted to lower $R$ as compare the $R_e$. 
Changes of $V^0_{nl}(R)$  with the quantum numbers $n,l$ can be formulated as follows: a 
position of minimum shifts to lower $R$ and a depth of minimum is increasing (a) with increasing $n$ for circulate orbits ($l=n-1$); 
(b) with decreasing of $l$ at fixed $n$.  

\begin{figure}[thb]  
\centering
    \includegraphics[width=0.48\textwidth]{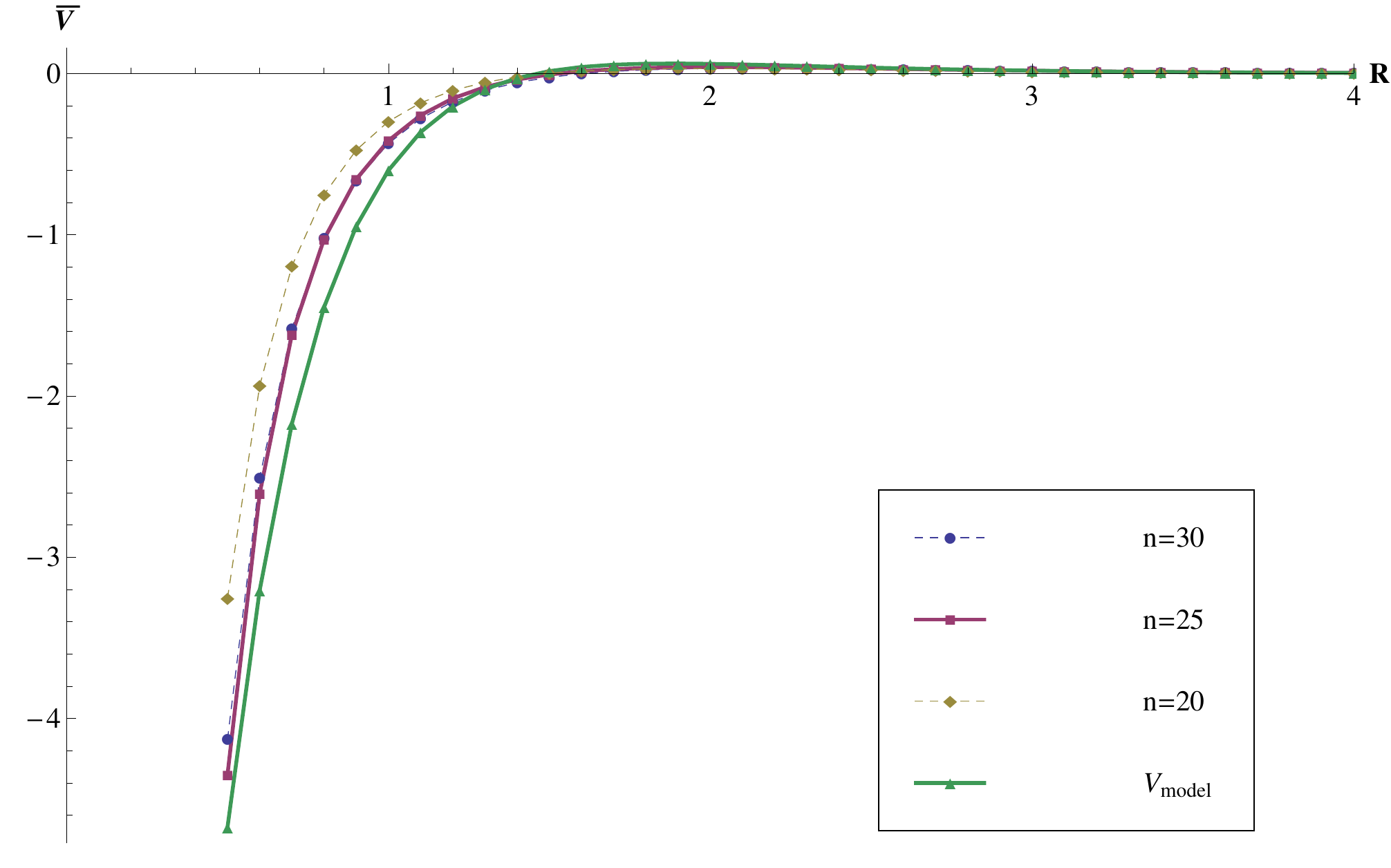}
		\put(-120,140){(a)}
	   \includegraphics[width=0.48\textwidth]{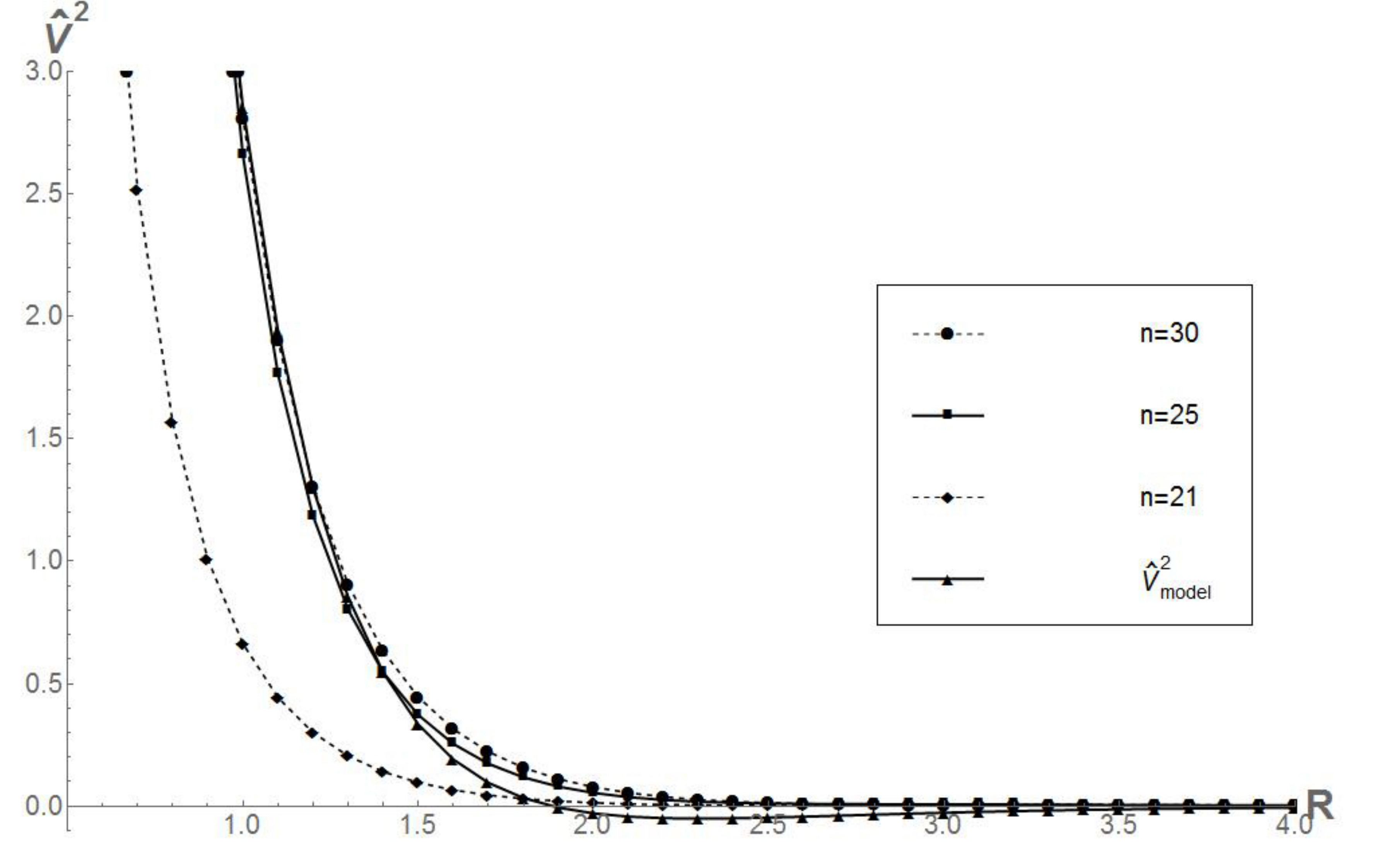}
		\put(-120,140){(b)}
\caption{Dependencies of $\tilde{V}^1_{nn-2,nn-1}(R)$ and $\tilde{V}^2_{nn-3,nn-1}(R)$ 
 on $R$ for circular orbits at different $n$ in comparison with the model values.}
    \label{fig7}
\end{figure}

Non-diagonal matrix elements \eqref{eq11} at $k=1,\,2$ have to be compared with the 
corresponding matrix elements of the model potential. The dependencies of the model matrix elements on quantum numbers can 
be separated out with factors
\begin{align}
 I_k(nl,nl') & = \langle R_{nl'}(r)|r^k|R_{nl}\rangle , \label{eq13} \\
 I_1(nl,nl-1)& =\frac{3}{2}n\sqrt{n^2-l^2}/(\mu Z), \label{eq14} \\
 I_2(nl,nl)& =\frac{1}{2}n^2[5n^2 + 1 - 3l(l+1)]/(\mu Z)^2 , \label{eq15}\\
 I_2(nl,nl-2)& =-\frac{5}{2}n^2\sqrt{(n^2 - l^2)[n^2-(l-1)^2]}/(\mu Z)^2. 
\label{eq16} 
\end{align}
In order to exclude effects of these factors, we introduce 'scaled' matrix elements
\begin{equation}
\tilde{V}^k_{nl',nl}(R)= V^k_{nl',nl}(R)/\left( 2^{-k}\xi_k I_k(nl,nl')\right)
 \label{eq17} 
\end{equation} 
that are independent on quantum numbers in the model, but can depend for the 
\emph{ab initio} potentials.
Radial dependencies of the dipole and quadruple terms in the model potential  
\eqref{eq1} are defined by simple factors $V_0^{\prime}(R)$ and $ R\left(R^{-1}V_0^{\prime}(R)
\right)^{\prime}$ that are independent on the quantum numbers. Fig. \ref{fig7} show 
dependencies of the 'scaled' \emph{ab initio} dipole and quadruple matrix elements for the circular orbits in comparison with the 
model. It is seen that the difference between the \emph{ab initio} and model matrix elements is appreciable at small and intermediate $R$ 
($R\lesssim 1.5$ for $k=1$, and $R\lesssim 2$ for $k=2$) but it is very small at higher $R$.

\begin{figure}[thb]  
\centering
    \includegraphics[width=0.7\textwidth]{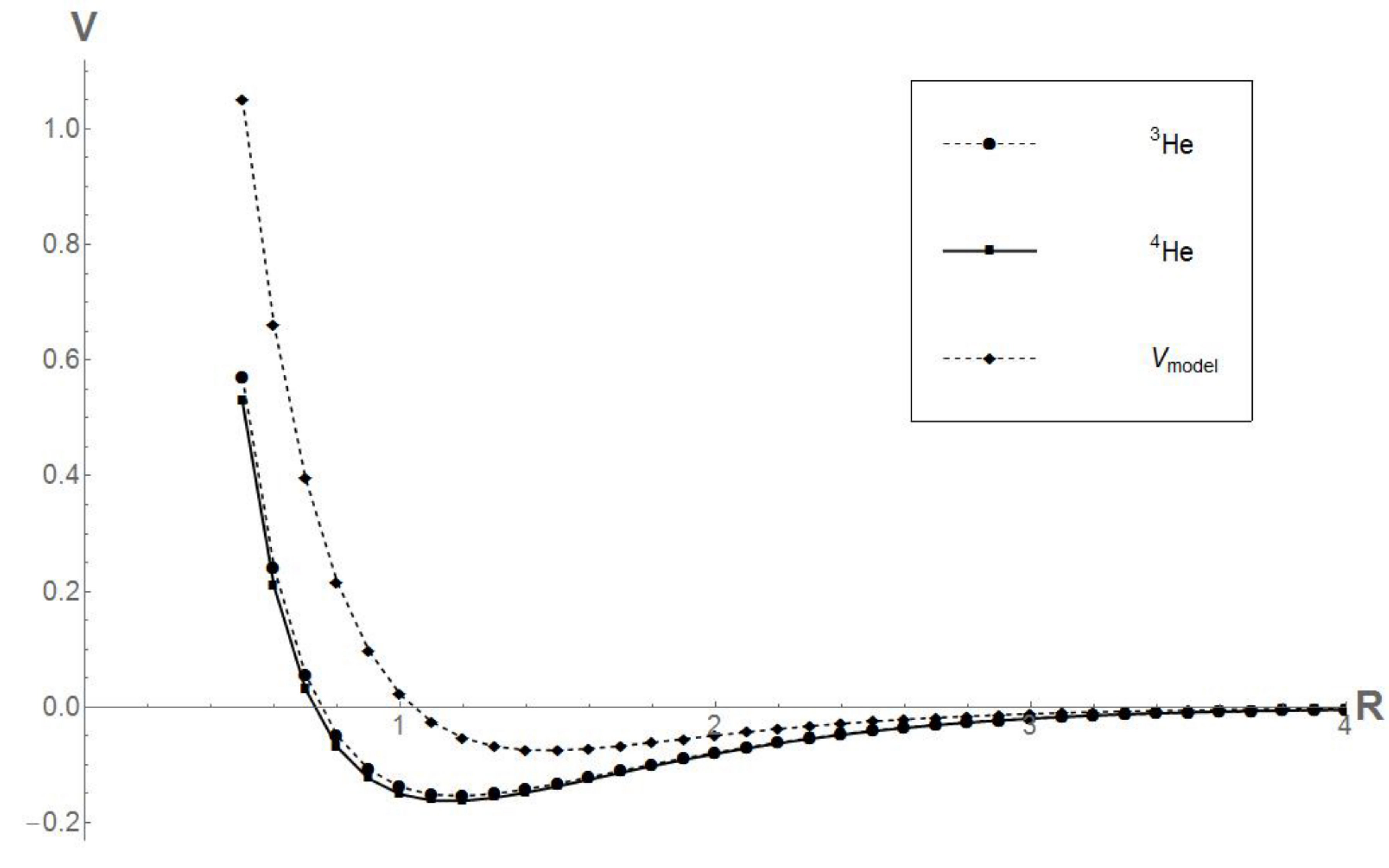}
	\caption{Monopole terms $V^0(R)$ for $^4\mathrm{He}$ and  $^3\mathrm{He}$ at $n=30,\,l
=n-1$ in comparison with the model potential $V_0(R)$.}
    \label{fig8}
\end{figure}

All results of the calculations shown above refer to the system with two $^4\mathrm{He}$ nuclei. 
We have done also the same calculations for the ($\mathrm{\bar{p}} - \mathrm{He}^{2+} - 
\mathrm{He}$)  system with $^3\mathrm{He}$ nuclei. Potential energy surfaces as well as the diagonal and non-
diagonal matrix elements for two isotopes are very close to each other.
 As an example we show in Fig. \ref{fig8} a comparison of monopole terms $V^0(R)$ for $^4\mathrm{He}$ and  $^3\mathrm{He}$ at $n=30,\,l=n-1$ together with the model potential $V_0(R)$. It is seen that \emph{ab initio} potentials for two isotopes are  practically indistinguishable, and the both differ markedly from the model potential. 
It should be noted, however, that in spite of the very close potentials, an isotope effect on dynamic characteristics (cross sections, transition rates, \emph{etc.}) can be appreciable due the difference of the masses. 

\begin{figure}[thb]  
\centering
    \includegraphics[width=0.9\textwidth]{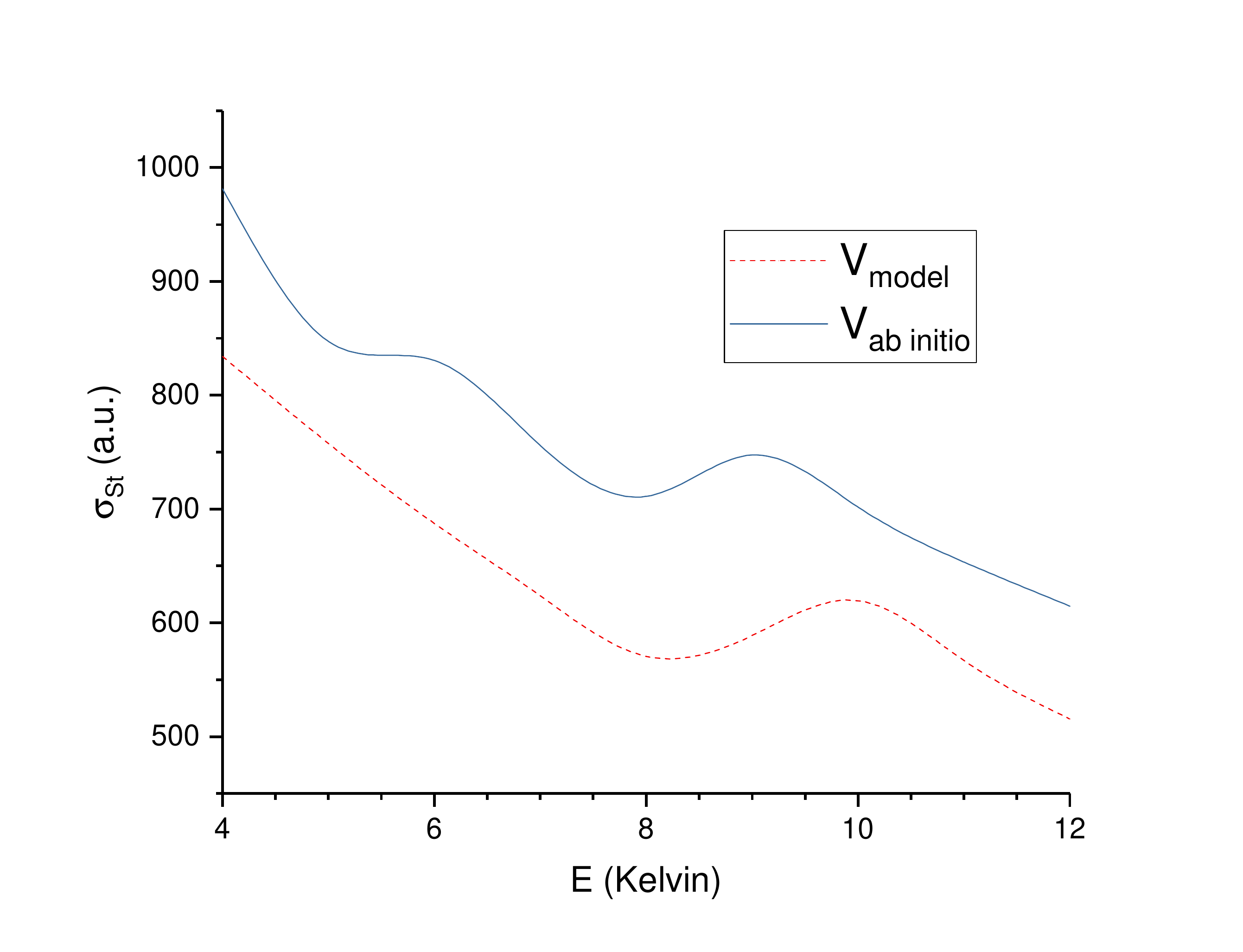}
	\caption{Summary cross sections of collisional Stark transition from $n=30,\,l=29$ to $l'<29$ in $(\bar{p}^4\mathrm{He}^{2+}$ obtained with PES and model potentials.}
    \label{fig9}
\end{figure}

We use the obtained PES potentials to calculate the cross sections of collisional Stark transition between the high states ($n\sim 30$) of $(\mathrm{\bar{p}He}^{2+})$ ion. The calculations of Stark cross sections were done by the 
same close coupling method as with the model potential\cite{ref12, ref13}. Fig. \ref{fig9} shows the results obtained with \emph{ab initio} and model potentials for the summary cross section of Stark transition from the state $n=30,\, l=29$ to $l'<29$ depending on energy. The model results contain a broad bump around 10 K that in fact is a superimposition of many resonances in different channels and partial waves \cite{ref12, ref13}. The cross section obtained with \emph{ab initio} potentials has a more complicated energy dependence (two splitted bumps shifted to lower energies), because the \emph{ab initio} potentials are deeper and depend on quantum number $l$ of output channels. In the considered energy region ($E\leq 12$ K) the summary Stark cross section obtained with the PES potentials exceeds the model results by 
15 - 20\%. 

\section{Conclusion} \label{sec5}
We have calculated the \emph{ab initio} potential energy surface for the 
($\mathrm{\bar{p}} - \mathrm{He}^{2+} - \mathrm{He}$) system in the framework of the restricted Hartree-Fock method with account of the electronic correlations within the second order perturbation method (MP2). Dependencies of the potential energy surface on $R$, $r$ and $\cos\theta$ in the most interesting region are considered. At $r\leq0.6,\, R\geq 1$ the angular dependence is rather weak that allows to use only low terms of multipole expansion. Dependencies on $R$ and $r$ are qualitatively similar to the model \cite{ref12, ref13}. Matrix of the potentials also similar to the model, however there are essential quantitative distinctions. Therefore the Stark cross sections calculated with the PES different markedly from the model. For the state $n=30,\,l=29$ in $(\bar{p}^4\mathrm{He}^{2+})$ the summary Stark cross section at 
$E\leq 12$ K obtained with PES potentials exceed the model results by 15 - 20\%. The more detailed results for the Stark cross sections as well as for transition rates averaged over termal motion and for the effective annihilation rates will be published separately.

\section*{Acknowledgements}
One of the authors (G.K.) thanks to E.Widmann and M.Hori for the interest to our work and useful discussions.


\begin{thebibliography}{99}\itemsep -1mm
\bibitem{ref1} T. Yamazaki, N. Morita, R.S. Hayano, E. Widmann and
J. Eades, Physics Reports \textbf{366}, 183 - 329 (2002).

\bibitem{ref2} R.S. Hayano, M. Hori, D. Horv\'{a}th and E. Widmann, Rep. Prog. Phys. \textbf{70},
1995 - 2065 (2007).
 
\bibitem{ref3} E. Widmann, J. Eades, T. Ishikawa, \emph{et al.},  Phys. Rev. Lett. \textbf{89},
243402 (2002).

\bibitem{ref4} T. Pask, D. Barna, A. Dax, \emph{et al.},  Phys. Letters B \textbf{678}, 55 - 59 (2009) 

\bibitem{ref5} S. Friedreich, D. Barna, F. Caspers, \emph{et al.},
J. Phys. B: At. Mol. Opt. Phys. \textbf{46}, 125003 (2013).

\bibitem{ref6} G.Ya. Korenman and S.N. Yudin,  J. Phys. B: At.
Mol. Opt. Phys. \textbf{39}, 1473 (2006).

\bibitem{ref7} S. Yudin and G. Korenman, Hyperfine Interactions \textbf{209}, 21
(2012).

\bibitem{ref8} M. Hori, J. Eades, R.S. Hayano, \emph{et al.}, Phys. Rev. Lett.
\textbf{94}, 063401 (2005).

\bibitem{ref9} G. Reifenr\"{o}ther, E. Klempt, and R. Landua, Phys. Lett. B  \textbf{203}, 9 (1988).

\bibitem{ref10} R. Landua and E. Klempt, Phys. Rev. Lett. \textbf{48}, 1722 (1982).

\bibitem{ref11} T.S. Jensen and V.E. Markushin, Eur. Phys. J. D \textbf{19}, 165 (2002).

\bibitem{ref12} G.Ya. Korenman and S.N. Yudin, J. of Phys.: Conf. Series \textbf{88},
012060 (2007).

\bibitem{ref13} G.Ya. Korenman and S.N. Yudin, Proc. of International Conf. on Muon Catalyzed
Fusion and Related Topics (MCF-07), Dubna, JINR, 2008, p. 191-198.

\bibitem{ref14} J.E. Russel, Phys. Rev. A \textbf{34}, 3865 (1986).

\bibitem{ref15} J. Cohen, Phys. Rev. A \textbf{25}, 1791 (1982).

\bibitem{ref16} D. Bakalov, B. Jeziorski, T. Korona, K. Szalewicz, and E. Tchoukova, Phys. Rev. Lett. \textbf{84}, 2350 (2000).

\bibitem{ref17} S. Sauge, P. Valiron, Chemical Physics \textbf{265}, 47 (2001).

\bibitem{ref18} S. Sauge, P. Valiron, Chemical Physics \textbf{283}, 433 (2002).

\bibitem{ref19} T.H. Dunning, Jr., J. Chem. Phys. \textbf{90}, 1007 (1989).

\bibitem{ref20} D.E. Woon and T.H. Dunning, Jr. J. Chem. Phys. \textbf{100}, 2975 (1994).

\bibitem{ref21} Extensible Computational Chemistry Environment Basis set Database,\\ Version 02/02/06 (EMSL), http://www.emsl.pnl.gov/forms/basisform.html. Current version of the database is at https://bse.pnl.gov/bse/portal.

\bibitem{ref22} A. Artemyev, A. Bibikov, V. Zayets, and I. Bodrenko,
  J. Chem. Phys. \textbf{123}, 024103 {2005}).

\bibitem{ref23} A.V. Nikolaev, I.V. Bodrenko, and E.V. Tkalya, Phys. Rev. A \textbf{77},
  012503 (2008).


\end{thebibliography}
\end{document}